\documentclass[structabstract]{aa}  
\usepackage{graphicx}
\usepackage{natbib}
\usepackage{epstopdf}
\usepackage{txfonts}

\begin{document}

\newcommand{\ohhd}{o-H$_2$D$^+$}
\newcommand{\hhdp}{H$_2$D$^+$}
\newcommand{\pddh}{p-D$_2$H$^+$}
\newcommand{\ddhp}{D$_2$H$^+$}

   \title{Extended emission of D$_2$H$^+$ in a prestellar core\thanks{Based on observations with the APEX telescope. APEX is a collaboration between the Max-Planck-Institut f\"ur Radioastronomie, the European Southern Observatory, and the Onsala Space Observatory}}

   \author{B. Parise
          \and 
   A. Belloche 
       \and 
   F. Du 
       \and
   R. G\"usten 
       \and
    K. M. Menten 
          }

   \institute{Max-Planck-Institut f\"ur Radioastronomie, Auf dem H\"ugel 69, 53121 Bonn, Germany\\
              \email{bparise@mpifr.de}
           }

   \date{Received ; accepted }

 
  \abstract
   { In the last years, the H$_2$D$^+$ and D$_2$H$^+$ molecules have gained great attention as probes of cold and depleted dense molecular cloud cores. These ions are at the basis of molecular deuterium fractionation, a common characteristic observed in star forming regions. H$_2$D$^+$ is now routinely observed, but the search for its isotopologue D$_2$H$^+$ is still difficult because of the high frequency of its ground para transition (692 GHz). }
   {We have observed molecular transitions of H$_2$D$^+$ and D$_2$H$^+$ in a cold prestellar core to characterize the roots of deuterium chemistry.}
   {Thanks to the sensitive multi-pixel CHAMP$^+$ receiver on the APEX telescope where the required excellent weather conditions are met, we not only successfully detect D$_2$H$^+$ in the H-MM1 prestellar core located in the L1688 cloud, but also obtain information on the spatial extent of its emission. We also detect H$_2$D$^+$ at 372\,GHz in the same source. We analyse these detections using a non-LTE radiative transfer code and a state-of-the-art spin-dependent chemical model.   }
   {This observation is the first secure detection of D$_2$H$^+$ in space. The emission is moreover extended over several pixels of the CHAMP$^+$ array, i.e. on a scale of at least 40$''$, corresponding to $\sim$\,4800\,AU. We derive column densities on the order of 10$^{12}$\,--\,10$^{13}$\,cm$^{-2}$ for both molecules in the LTE approximation depending on the assumed temperature, and up to two orders of magnitude higher based on a non-LTE analysis. }  
{Our modeling suggests that the level of CO depletion must be extremely high ($>$10, and even $>$100 if the temperature of the core is around 10\,K) at the core center, in contradiction with CO depletion levels directly measured in other cores. Observation of the H$_2$D$^+$ spatial distribution and direct measurement of the CO depletion in H-MM1 will be essential to confirm if present chemical models investigating the basis of deuterium fractionation of molecules need to be revised. }
   \maketitle
%

\section{Introduction}

Although deuterium is a trace element (D/H $\sim$10$^{-5}$), the study of deuterium fractionation in molecules is a very vivid field of research. The main astrochemical interest derives from the fact that the understanding of fractionation processes may ultimately help to understand the formation of the related hydrogenated species. As an example, the first detection of a doubly-deuterated molecule, D$_2$CO, by \citet{Turner90} was interpreted as the hint that grain chemistry may be at work for synthesizing formaldehyde. However, the impact of deuterium fractionation studies rapidly outpassed the mere astrochemical interest. Since deuterated molecules are formed preferentially at low temperatures and high densities, they became valuable probes for studying, e.g., the kinematics of cold cloud cores harbouring the earliest stages of star formation \citep{vanderTak05}. 

The basic process transferring deuterium from the HD molecule to other molecules is through ion-molecule reactions involving H$_3^+$, CH$_3^+$, and C$_2$H$_2^+$ \citep[see][and references therein]{Parise09}.
The route involving H$_3^+$ is the dominant one at very low temperatures ($T$\,$<$\,20\,K), whereas the importance of CH$_3^+$ takes over at slightly higher temperatures (20\,$<$\,$T$\,$<$\,40\,K) due to its higher endothermicity \citep{Turner01,Roueff07, Parise09}. H$_2$D$^+$ is thus a very good probe of very cold objects, and has been extensively used in the last years to probe gas in which  CO is strongly depleted, i.e., frozen out on dust grain surfaces \citep[e.g.][]{Caselli03}. The deuteration level is also a decisive argument in the study of the evolutionary state of early phases of star formation. The deuterium fractionation is believed to decrease once the central protostar starts to heat its envelope \citep[e. g., ][]{Emprechtinger09}. For instance, partly based on molecular deuteration, \citet{Belloche06} concluded that the dense core 
Cha-MMS1 in the Chamaeleon~I molecular cloud is likely at the stage of the first hydrostatic core when H$_2$ has not been dissociated yet.

The study of several isotopologues of methanol toward the low-mass protostar IRAS16293-2422 \citep{Parise02, Parise04} showed that grain chemistry models require a very high atomic D/H ratio accreting on the grains in order to explain the high CHD$_2$OH and CD$_3$OH abundances. At the time, gas-phase models could not reproduce such a high atomic D/H ratio, but a decisive step forward was done with the inclusion of D$_2$H$^+$ and D$_3^+$ in the reaction networks \citep{Roberts03}. Since then, D$_2$H$^+$  was searched for through observation of its ground para state line. The frequency of this line was initially published to be 691.660440 GHz \citep[$\pm$\,19\,kHz (1$\sigma$),][]{Hirao03} and was later revised to 691.660483 GHz \citep[$\pm$\,20\,kHz (1$\sigma$),][]{Amano05}. These astronomical searches are extremely difficult because of the high frequency of the molecular line, for which excellent instruments and weather conditions are required. Using the Caltech Submillimeter Observatory 10.4\,m telescope (CSO), \citet{Vastel04} detected a $\sim$3.3$\sigma$  line (in peak) toward the Ophiuchus dense core 16293E (4.4$\sigma$ in integrated intensity). This detection is to date the only published one, and its low signal-to-noise ratio as well as the velocity shift between H$_2$D$^+$ and D$_2$H$^+$ \citep[0.23\,km s$^{-1}$ after remeasurement of \ohhd~and \pddh~frequencies,][]{Amano05}   certainly require a confirmation of the detection by improving the observation sensitivity and frequency calibration precision (which was $\sim$\,0.1\,km\,s$^{-1}$ in the observations of \citet{Vastel04}). We present here the first secure detection of D$_2$H$^+$ toward the prestellar core H-MM1 in L1688, the main molecular cloud in the Ophiuchus star forming region.

\section{Observations}

\subsection{The source}

\begin{figure}[!h]
\includegraphics[width=7.0cm,angle=-90]{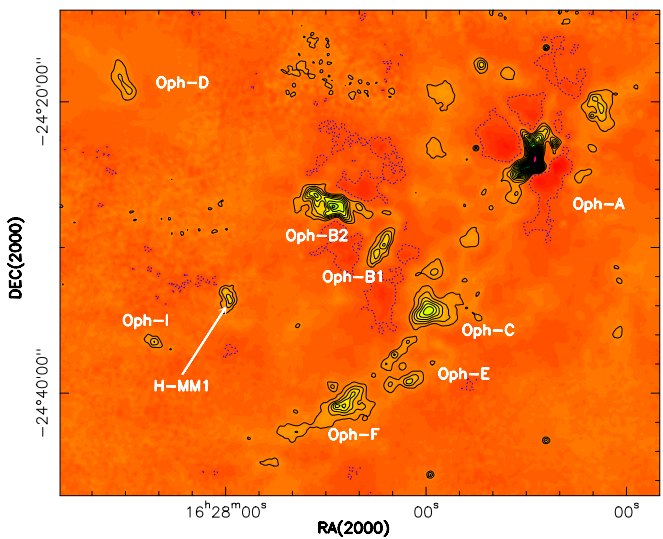}
\caption{The L1688 region, as mapped by SCUBA (850 $\mu$m), from the SCUBA legacy catalogues (see text). The contours are in steps of 3$\sigma$ (=0.18 Jy/beam). The beamsize is 14$''$ (HPBW). }
\label{rho_oph}
\end{figure}

The L1688 cloud (the main cloud in the Ophiuchus dark cloud) has been mapped both in molecular tracers \citep[e.g.][]{Loren90} and in continuum emission \citep[e.g. at 1.3\,mm by][]{Motte98}. These early mapping studies revealed different areas of high column density, named Oph-A to Oph-G. Using the Submillimeter Common User Bolometer Array (SCUBA) at the 15\,m James-Clerck-Maxwell Telescope (JCMT), \citet{Johnstone04} covered a larger region, leading to the detection of two new dense cores, that they named Oph-H and Oph-I, following the same nomenclature. Oph-H contains a single continuum peak,  H-MM1.
The distance of the Ophiuchus region has been recently accurately determined to 120\,pc \citep{Loinard08}.

We present in Fig. \ref{rho_oph} a part of the SCUBA map of the  Ophiuchus region, as available from the SCUBA Legacy catalogues\footnote{http://www2.cadc-ccda.hia-iha.nrc-cnrc.gc.ca/community/ scubalegacy/} \citep[data: scuba\_F\_353d1\_16d8\_850um.emi.fits,][]{diFrancesco08}, showing the relative position of all dense cores. In this paper, we present observations toward the H-MM1 core, at coordinates $\alpha _{2000}$=16$^h$27$^m$58.3$^s$ and $\delta_{2000}$=$-$24$^\circ$33$'$42.0$''$.

H-MM1 corresponds to the Bolo25 source of the 1.1\,mm BOLOCAM (CSO) map observed by \citet{Young06}. 
The FWHM beam size of these observations is 31$''$. Assuming a temperature of 10\,K  and a dust opacity\footnote{This dust opacity assumes a gas-to-dust mass ratio of 100, and is expressed per dust+gas mass unit} $\kappa_{\nu}$ = 0.0114 cm$^2$\,g$^{-1}$ at 1.12\,mm these authors infer a total mass of 0.24\,M$_{\odot}$ in 40$''$, and measure a source FWHM size of 47$''$$\times$\,56$''$. They measure a peak visual extinction,  $A_{\mathrm{V}}$,  of 23 mag, and a mean density of 5$\times$10$^5$\,cm$^{-3}$.  

\citet{Khanzadyan04} and \citet{Stanke06} carried out an unbiased search for protostellar signatures in the $\rho$-Oph cloud. The closest source they identify in their continuum maps is MMS38 (at 9$''$ from our position), which they classify as starless. No optical/NIR counterpart (2MASS) or sign of outflow are found in this source. 

The region was also mapped with the \textit{Spitzer} Space telescope, in the frame of the \textit{Cores to Disks} legacy project (c2d). We downloaded the corresponding data from the Spitzer website\footnote{http:$//$data.spitzer.caltech.edu/popular/c2d/20071101\_enhanced\_v1/ oph/}. No source is detected in any of the \textit{Spitzer} bands, especially at 70 $\mu$m, which is a good tracer of embedded protostars \citep[see][]{Dunham08}. H-MM1 thus appears to be a dense prestellar core.

\subsection{APEX line observations}

Using the APEX2a single-pixel and CHAMP$^+$ array receivers at the Atacama Pathfinder Experiment telescope (APEX), we observed H$_2$D$^+$ (372 GHz) and D$_2$H$^+$ (692 GHz) toward H-MM1 at the position given in Sect. 2.1. We used in each observation the position-switch mode, with the off-position at (0$''$, 300$''$) equatorial offsets. Calibration was done by measuring regularly the sky brightness and cold and hot loads in the cabin.

\subsubsection{H$_2$D$^+$}

Using the APEX2a receiver \citep{Risacher06}, we observed the ground-state \ohhd~line toward H-MM1. The frequency of this line is 372.421385 GHz \citep{Amano05}. Observations took place on June 29th, 2007, under very good weather conditions (precipitable water vapor PWV=0.24\,mm). The system temperature was 250\,K. The backend, a Fast Fourier Transform Spectrometer (FFTS), was used with 8192 channels, yielding a 0.0983\,km s$^{-1}$ channel separation. The effective frequency resolution of the FFTS in this configuration is 159\,kHz \citep[][and B. Klein, {\it priv. comm.}]{Klein06}, i.e. 0.13\,km/s at 372 GHz.

The beam efficiency of APEX at 372\,GHz was measured to be 0.73 for compact sources \citep[as measured on Mars and Jupiter, which were respectively 8$''$ and 32-35$''$ at that time,][]{Guesten06SPIE}.
As the D$_2$H$^+$ emission seems extended on a scale comparable to the footprint of CHAMP$^+$ ($\sim$40$''$, see Figure \ref{raster}), H$_2$D$^+$ is likely to also be extended on that scale. We therefore adopt the $T_{mb}$  scale for our analysis, using the beam efficiency of 0.73. The forward efficiency of the telescope is 0.95 at this frequency.

We reached a rms noise level of 72\,mK ($T_a^*$ scale) at the initial resolution (0.13\,km\,s$^{-1}$).

\subsubsection{D$_2$H$^+$}

\begin{figure}
\includegraphics[width=8cm,angle=-90]{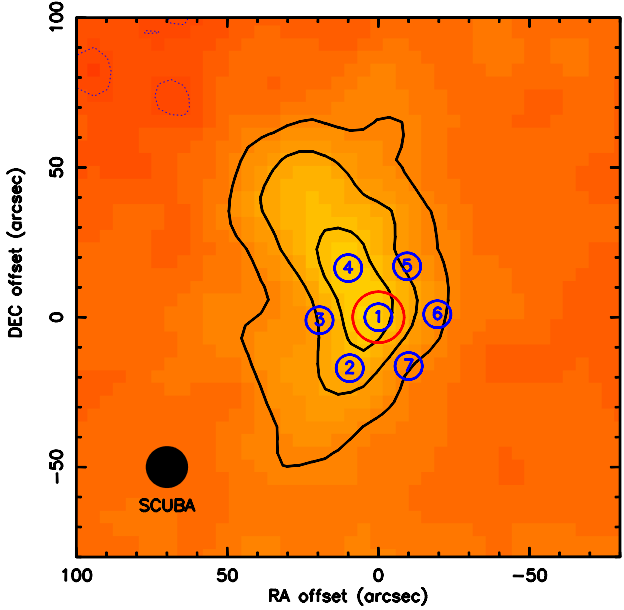}
\caption{850 $\mu$m continuum emission of the H-MM1 core. The map is centered on the position of H-MM1 (see text). The contours are in steps of 3$\sigma$ (=0.18 Jy/beam). The blue numbers show the different  pixels of the CHAMP$^+$ low-frequency array. The blue circles show the D$_2$H$^+$ beams. In red, the H$_2$D$^+$ beam. }
\label{footprint}
\end{figure}

Using the CHAMP$^+$ array receiver on the APEX telescope, we targeted the prestellar core H-MM1.
The observations took place 
in June and August 2009, 
under good to very good weather conditions (most of the time with PWV\,$<$\,0.45mm). The typical system temperature in the central pixel of the array varied between 800 and 1400\,K. 
Figure \ref{footprint} presents the overlay of the CHAMP$^+$ footprint on the 850 $\mu$m continuum emission. Each CHAMP$^+$ pixel was tuned to the CO(6-5) frequency (691.4730763 GHz), which is separated by 187.4\,MHz from the D$_2$H$^+$ frequency. We adopt here the most recent measurement of \citet{Amano05} of the D$_2$H$^+$ line frequency: 691.660483 GHz. The beam size is 9$''$. The CHAMP$^+$ AFFTS was used with a channel separation of 183 kHz (0.079 km/s). The effective resolution is 212\,kHz, i.e. 0.092\,km/s. 

The beam efficiency at 692 GHz was measured to be 0.52 on Jupiter (angular size 47.3$''$ at that time). Because the D$_2$H$^+$ emission is observed to be extended (see below) on a typical size of 40$''$, we adopt the beam efficiency of 0.52. The forward efficiency is 0.95 at this frequency.\\

We reached an rms noise level of 34\,mK ($T_a^*$ scale) at the initial resolution (0.09\,km\,s$^{-1}$).

\subsection{Continuum data}

We retrieved the 850$\mu$m map of the Ophiuchus region from the SCUBA Legacy catalogues \citep[data: scuba\_F\_353d1\_16d8\_850um.emi.fits][]{diFrancesco08}. The region covering the L1688 cloud is shown in Fig. \ref{rho_oph}, and was discussed by \citet{Jorgensen08}.

 The beam of SCUBA at this frequency is 14$''$. The local rms in the H-MM1 region of the map is found to be 60 mJy/beam.

\begin{figure}
\includegraphics[width=10 cm,angle=-90]{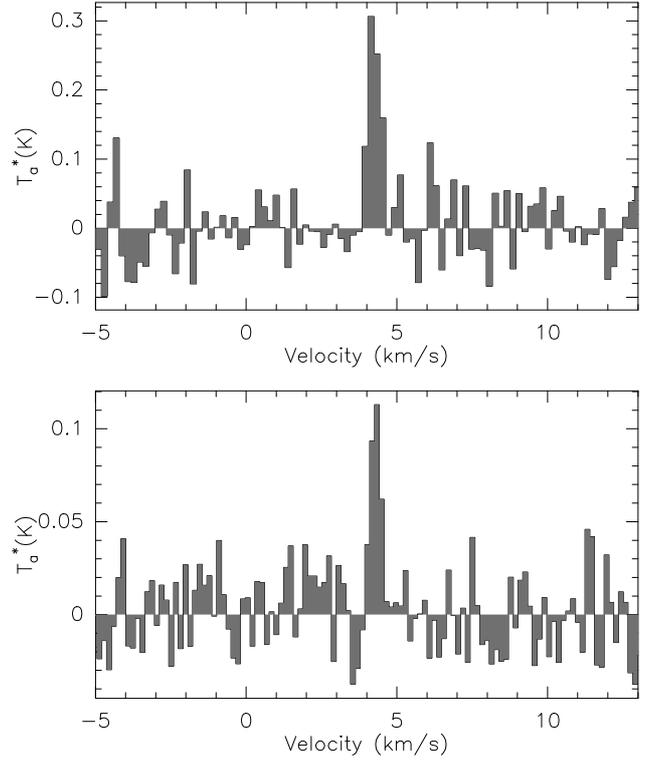}
\caption{H$_2$D$^+$ (upper panel) and D$_2$H$^+$ (lower panel) toward the central position of H-MM1.}
\label{overlay}
\end{figure}

\begin{figure}
\includegraphics[width=7.5cm,angle=-90]{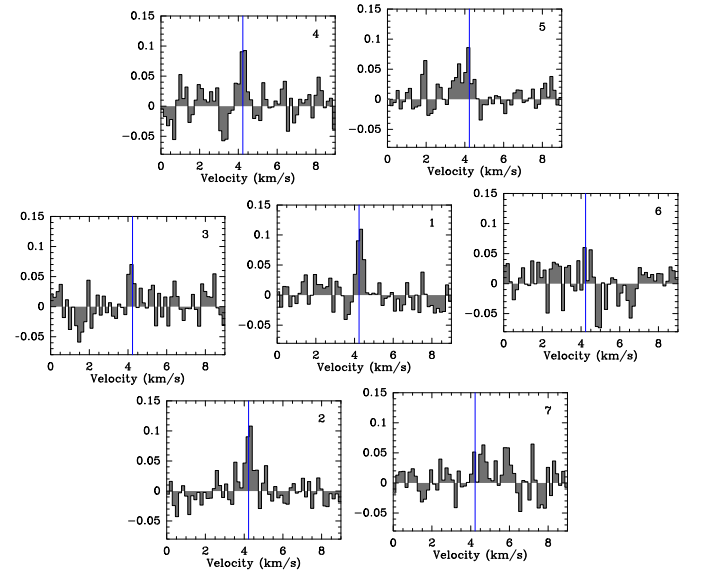}
\caption{D$_2$H$^+$ observations. Each panel shows a pixel of CHAMP$^+$ (as displayed on Fig. \ref{footprint}). The vertical line shows the v$_{\rm lsr}$ of the source as measured with the \ohhd~line (4.23\,km\,s$^{-1}$). The y-axis is in Kelvin, on the T$_a^*$ scale. }
\label{raster}
\end{figure}

\section{Results}

\begin{table*}[!ht]
\caption{Observational results.}
\label{obsres}
\begin{tabular}{lccccccccc}
\noalign{\smallskip}
\hline
\hline
\noalign{\smallskip}
Species   &    Position   & Pixel  & $\int T_{a}^*$ dv &  v$_{lsr}$  & $\Delta$\,v   & $T_{a,peak}^*$ &  [$\Delta$\,v]$_{deconv}$ & $S^\star$  &  $\int T_{a}^*$ dv /  $S$\\ 
 & EQ offsets& & (mK km s$^{-1}$)  & (km s$^{-1}$) & (km s$^{-1}$) & (K) & (km s$^{-1}$) & (mJy/beam) & {\footnotesize (K km s$^{-1}$ / [Jy/beam]) }\\
\noalign{\smallskip}
\hline
\noalign{\smallskip}
o-H$_2$D$^+$  & (0$''$,\,0$''$)  &1& 180$\pm$20  &  4.23$\pm$0.04 &  0.51$\pm$0.07 & 0.33 & 0.49 \\
\noalign{\smallskip}
\hline
\noalign{\smallskip}
p-D$_2$H$^+$  & (0$''$,\,0$''$)  & 1& 47$\pm$8 &  4.30$\pm$0.03 &  0.35$\pm$0.06 & 0.13  &  0.34 &  628$\pm$60   & 0.076$\pm$0.015  \\
& (9.5$''$,\,$-$17$''$)  &2&  49$\pm$8 &  4.26$\pm$0.04 &  0.42$\pm$0.09 & 0.11 &   0.41 & 468$\pm$60   &  0.106$\pm$0.022 \\
  & (19.5$''$,\,$-$1$''$)  & 3& 27$\pm$10   &  4.16$\pm$0.05 &  0.31$\pm$0.10 & 0.08 & 0.30 &  404$\pm$60 & 0.068$\pm$0.027 \\  
  & (10$''$,\,16$''$)  & 4& 44$\pm$11 &  4.23$\pm$0.05 &  0.41$\pm$0.12 & 0.10 & 0.40 &   616$\pm$60  & 0.072$\pm$0.019 \\
 & ($-$9.5$''$,\,17.0$''$) & 5& $<$ 30 & -- & -- & -- & -- &   343$\pm$60  & $<$ 0.087 \\
 & ($-$19.5$''$,\,1$''$) & 6&$<$ 20  & -- & -- & --  & -- &  211$\pm$60 & $<$ 0.095  \\
  & ($-$10$''$,\,$-$16.3$''$) &7& $<$ 20  & -- & -- & -- & -- &  179$\pm$60 &  $<$ 0.112  \\
\noalign{\smallskip}
\hline
\noalign{\smallskip}
\end{tabular}\\
$^\star$\,$S$ is the peak flux density in the 850\,$\mu$m SCUBA map at the center of each CHAMP$^+$ pixel. It is in units of mJy/14$''-$JCMT beam. The noise in the map is 60\,mJy/beam.
The error bars are all 1$\sigma$, and the upper limits 3$\sigma$.
\end{table*}

The main result of this work is the first unambiguous detection of p-D$_2$H$^+$ toward a prestellar core. 
Figure \ref{overlay} presents the H$_2$D$^+$ and D$_2$H$^+$ emission at the (0$''$, 0$''$) position. The agreement of the 
LSR velocities, v$_{lsr}$, for both lines is remarkable (see Table~\ref{obsres}) and further strengthens the assignment of the lines. Figure \ref{raster} presents the D$_2$H$^+$ observation on each of the CHAMP$^+$ pixels. The noise level is different from pixel to pixel, due to the different performances of the off-axis pixels. The line is clearly detected in at least two adjacent pixels (pixels 2 and 4). There is also a tentative detection in pixel 3, with a lower intensity. The D$_2$H$^+$ emission is thus extended on the scale of the CHAMP$^+$ footprint ($\sim$40$''$), corresponding to a physical size of $\sim$\,4800\,AU at the distance of 120\,pc. The observational results are listed in Table~\ref{obsres}.

We note that the detection of p-D$_2$H$^+$ in H-MM1 does not imply that it is a peculiar core. None of the  searches for p-D$_2$H$^+$ published in \citet{Caselli08} towards five prestellar cores have reached upper limits which are deep enough to discard the detection of the line with the same peak temperature in those cores.

\subsection{Linewidths}
\label{linewidth}

The measured linewidths are listed in Table \ref{obsres}. Because the lines are not very broad compared to the frequency resolution of the FFTS, we also computed deconvolved linewidths to correct for instrumental broadening (column 7).

The expected thermal linewidth for a molecule of mass m is:
$${\rm   \Delta v_{therm,FWHM} = 2 \times \sqrt{2\,{\rm ln} 2} \times \sqrt{\frac{kT}{m}}  }$$

For \hhdp, ${ \rm \Delta v_{therm} = 0.107 \times \sqrt{T}}$\,km/s. For \ddhp, ${\rm \Delta v_{therm} = 0.0955 \times \sqrt{T}}$\,km/s, with $T$ in Kelvin. 

The small linewidth of D$_2$H$^+$ on the (0$''$, 0$''$) position constrains the kinetic temperature to $T$\,$<$\,13\,K.  
For this upper limit value of the temperature, the rather large H$_2$D$^+$ linewidth points to non-thermal velocity broadening of 0.3\,km/s.
The different non-thermal velocity contribution to both lines might be a sign that the two molecules do not trace exactly the same gas. The larger non-thermal velocity broadening of \ohhd~may imply that \ohhd~is present in a more diffuse and turbulent gas than \pddh, as also supported by chemical models, which predict that the \pddh /\,\ohhd~ratio raises with density (see Figure \ref{obs_mod}). In this case, the \pddh /\,\ohhd~ratio that we will derive in the densest gas will likely be a lower limit.  

\subsection{Correlation with continuum emission}

The continuum emission at the position of  the individual pixels of CHAMP$^+$ is listed in Table \ref{obsres}. 
The positions where \pddh~is detected correspond to the positions where the continuum emission is the strongest. The lower tentative detection in pixel 3 is also consistent with the continuum being weaker at this position. The correlation between the \pddh~and the continuum emission is quite good. This implies that the \pddh~emission comes from the densest regions of the core, in agreement with the narrow width of the line.

\subsection{Column densities}

\subsubsection{LTE assumption}

The critical densities for the o-H$_2$D$^+$ and p-D$_2$H$^+$ transitions are, respectively, 1.3$\times$10$^5$ and 5.6$\times$10$^5$ cm$^{-3}$ \citep{Hugo09}. Although the density of the dense core may be of the same order or only marginally higher than these critical densities, we start with an LTE analysis as a first approximation. We will consider a non-LTE approach in the  following section.

We consider the ortho and para species of each molecule as different species, that do not interconvert via collisions with H$_2$. We thus compute separately the partition function of ortho and para species, from the transitions given in the Cologne Database of Molecular Spectroscopy \citep[CDMS, ][]{Muller05}. The partition functions are computed as $$ Q(T) = \sum_{levels} g_i g_r e^{(-E / kT)}$$ where $g_i$ is the spin degeneracy (1 and 3 for para- and ortho- \hhdp\ respectively, and 1 and 2 for para- and ortho- \ddhp) and g$_r$ the rotational degeneracy (g$_r$ = 2J + 1). Note that the energies of the levels were all expressed relative to the ground level 0$_{00}$ of each molecule. 
These partition functions are presented in Table~\ref{partitionFunction}.

\begin{table}[!h]
\caption{Partition functions of  \ohhd and \pddh at different temperatures.}
\label{partitionFunction}
\begin{center}
\begin{tabular}{ccc}
\noalign{\smallskip}
\hline
\hline
\noalign{\smallskip}
Temperature  &  Q(\ohhd) & Q(\pddh) \\
\tiny{(K)}  & & \\
\noalign{\smallskip}
\hline
\noalign{\smallskip}
7 &  4.2$\times$10$^{-5}$ &  2.3$\times$10$^{-3}$  \\
8 &  2.0$\times$10$^{-4}$ &  5.7$\times$10$^{-3}$  \\
9 &  6.9$\times$10$^{-4}$ &  1.2$\times$10$^{-2}$  \\
10 & 1.8$\times$10$^{-3}$ &  2.0$\times$10$^{-2}$  \\
11 & 4.2$\times$10$^{-3}$ &  3.3$\times$10$^{-2}$   \\
12 & 8.2$\times$10$^{-3}$ &  4.8$\times$10$^{-2}$  \\
13 & 1.5$\times$10$^{-2}$ &  6.8$\times$10$^{-2}$\\
\noalign{\smallskip}
\hline
\noalign{\smallskip}
\end{tabular}
\end{center}
\end{table}

\begin{table*}[!ht]
\caption{Line opacities and column densities on the (0$''$, 0$''$) position in the LTE hypothesis.}
\label{tabcoldens}
\begin{center}
\begin{tabular}{cccccc}
\noalign{\smallskip}
\hline
\hline
\noalign{\smallskip}
$T_{ex}$  &  $\tau$(\ohhd) & $N$(\ohhd) & $\tau$(\pddh) & $N$(\pddh) & $N$(\pddh)/$N$(\ohhd) \\
\tiny{(K)}  & & \tiny{(cm$^{-2}$)} & & \tiny{(cm$^{-2}$)} & \\
\noalign{\smallskip}
\hline
\noalign{\smallskip}
7 & 0.35 & 9.3$\times$10$^{12}$ & 1.7 & 4.1$\times$10$^{13}$ &  4.4 \\
8 & 0.23 &  6.5$\times$10$^{12}$&  0.59 &  1.5$\times$10$^{13}$  & 2.2 \\
9 & 0.17 &  5.1$\times$10$^{12}$ & 0.33 & 8.2$\times$10$^{12}$  & 1.6 \\
10 & 0.13 & 4.2$\times$10$^{12}$ & 0.21 &  5.4$\times$10$^{12}$ & 1.3 \\
11 & 0.11 & 3.6$\times$10$^{12}$&  0.15 & 3.9$\times$10$^{12}$   & 1.1\\
12 & 0.088 & 3.2$\times$10$^{12}$ & 0.11&  3.0$\times$10$^{12}$ & 0.96\\
13 & 0.076 & 2.9$\times$10$^{12}$ & 0.089 & 2.5$\times$10$^{12}$ &  0.86 \\
\noalign{\smallskip}
\hline
\noalign{\smallskip}
\end{tabular}\\
\end{center}
\end{table*}

The column density in the upper level of the transition 
can be directly derived from the measure of the opacity. 

$${  N_{\rm u}=\frac{1}{h B_{ul}} \frac{1}{e^{h\nu/kT_{ex}}-1}  ~\tau_{peak} \Delta {\rm v} }$$

The opacity of the transition, $\tau$, is given by:

\begin{eqnarray}
T_{\rm mb} = [J(T_{\rm ex}) - J(T_{bg})] [1 - e^{-\tau}] 
 \end{eqnarray}

where $J(T) = \frac{h\nu/k}{exp(h\nu/kT)-1} $

Assuming LTE (i.e. $T_{\rm ex}$ = $T_{\rm kin}$), we can then derive the total column density of the species: 

$$ { N_{tot}=\frac{8 \pi \nu^3}{c^3} \frac{Q(T_{ex})}{g_u A_{ul}} \frac{e^{E_u/ kT_{ex}}}{e^{h\nu/kT_{ex}}-1} ~\tau_{peak} \Delta {\rm v}  }$$

We used the A$_{ul}$ of \citet{Ramanlal04}. For both transitions, they are 11\% different from the value quoted in the CDMS database. We used the accurate frequencies derived by \citet{Amano05}.

We note that equation (1) implies that $T$$_{\rm mb}$\,$\le$\,J($T$$_{\rm ex})$. This condition for the 692\,GHz line of D$_2$H$^+$ translates into J($T_{\rm ex}$)\,$\ge$\,0.25\,K, i.e. $T$$_{\rm ex}$\,$\ge$\,7\,K. 
This condition as well as the previous linewidth argument allow to restrict the kinetic temperature to the range 7--13\,K.

The results are presented in Table \ref{tabcoldens}. Depending on the assumed temperature, the \pddh / \ohhd~ratio varies in the range 0.86 -- 4.4. 

\subsubsection{Out-of-equilibrium radiative transfer}
\label{outLTE}

\begin{figure}[!ht]
\includegraphics[width=9cm]{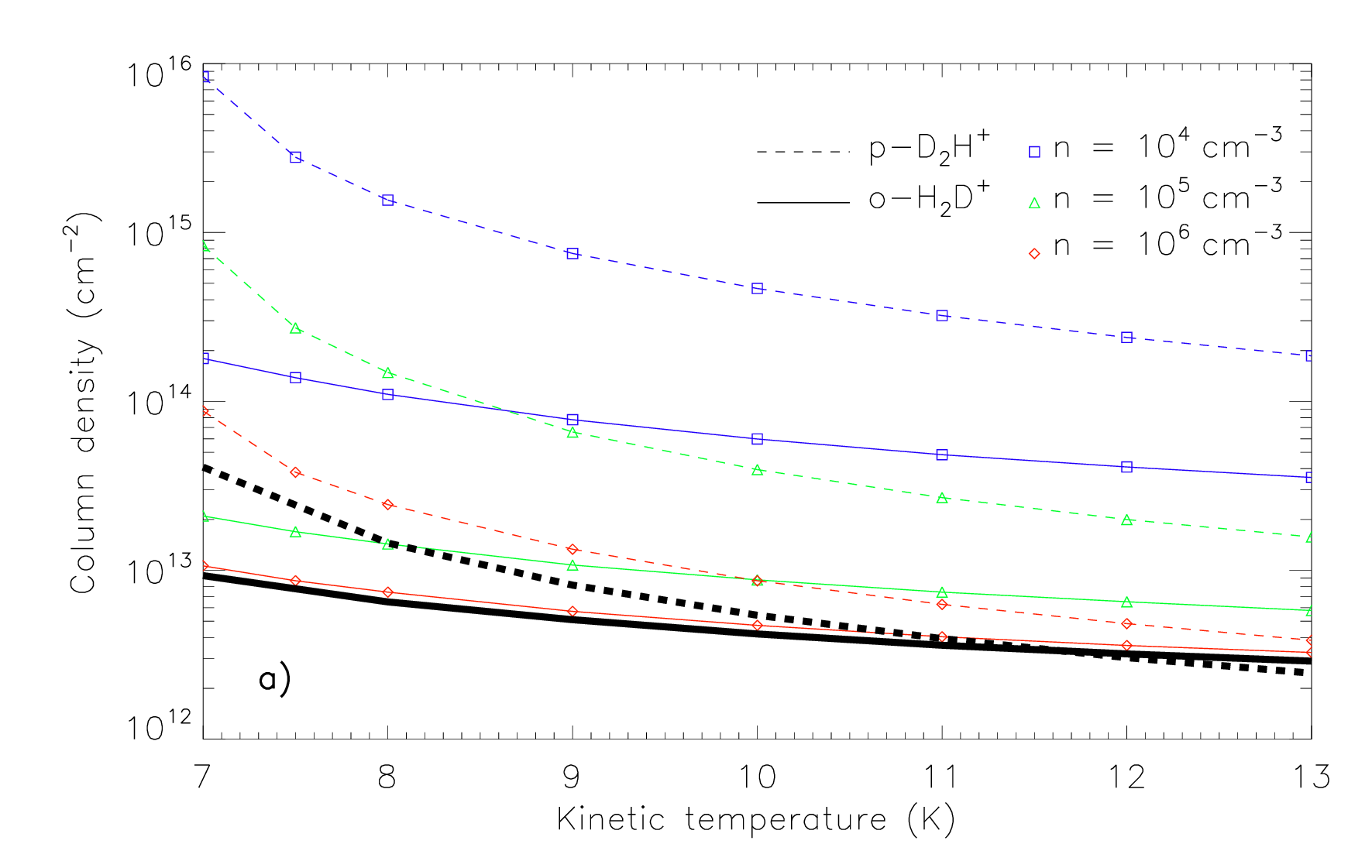}
\includegraphics[width=9cm]{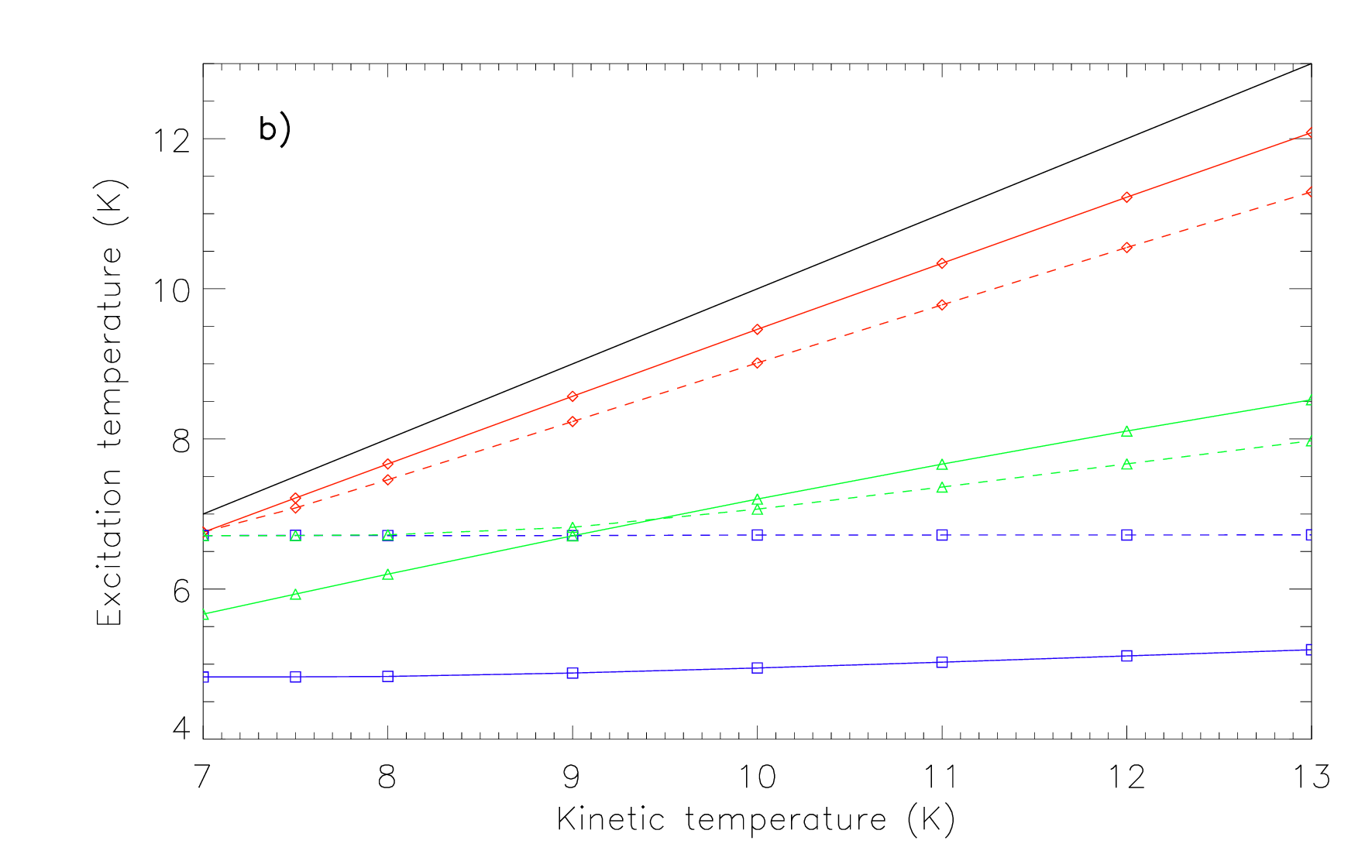}
\caption{{\bf a)} Column densities at the (0,0) position, computed from the non-equilibrium radiative transfer assuming the beam filling factor for both molecules is 1, as a function of the physical conditions (kinetic temperature, density). The thick black lines show the column densities computed in the LTE assumption. {\bf b)} Excitation temperature in the same conditions and color conventions as Figure a. The black line represents $T_{\rm ex}$\,=\,$T_{\rm kin}$. }
\label{coldens}
\end{figure}

Thanks to the recent computation of all elastic and non-elastic collision rate coefficients by \citet{Hugo09}, we can derive the column density using an out-of-equilibrium radiative transfer model. 
We used the non-LTE radiative transfer code RADEX \citep{vanderTak07}, in the isothermal sphere geometry, with the escape probability as a function of the opacity as follows: 

$$\beta = \frac{1.5}{\tau}[1- \frac{2}{\tau^2} +(\frac{2}{\tau}+ \frac{2}{\tau^2}) e^{-\tau}] $$

\begin{figure}
\includegraphics[width=10.5cm]{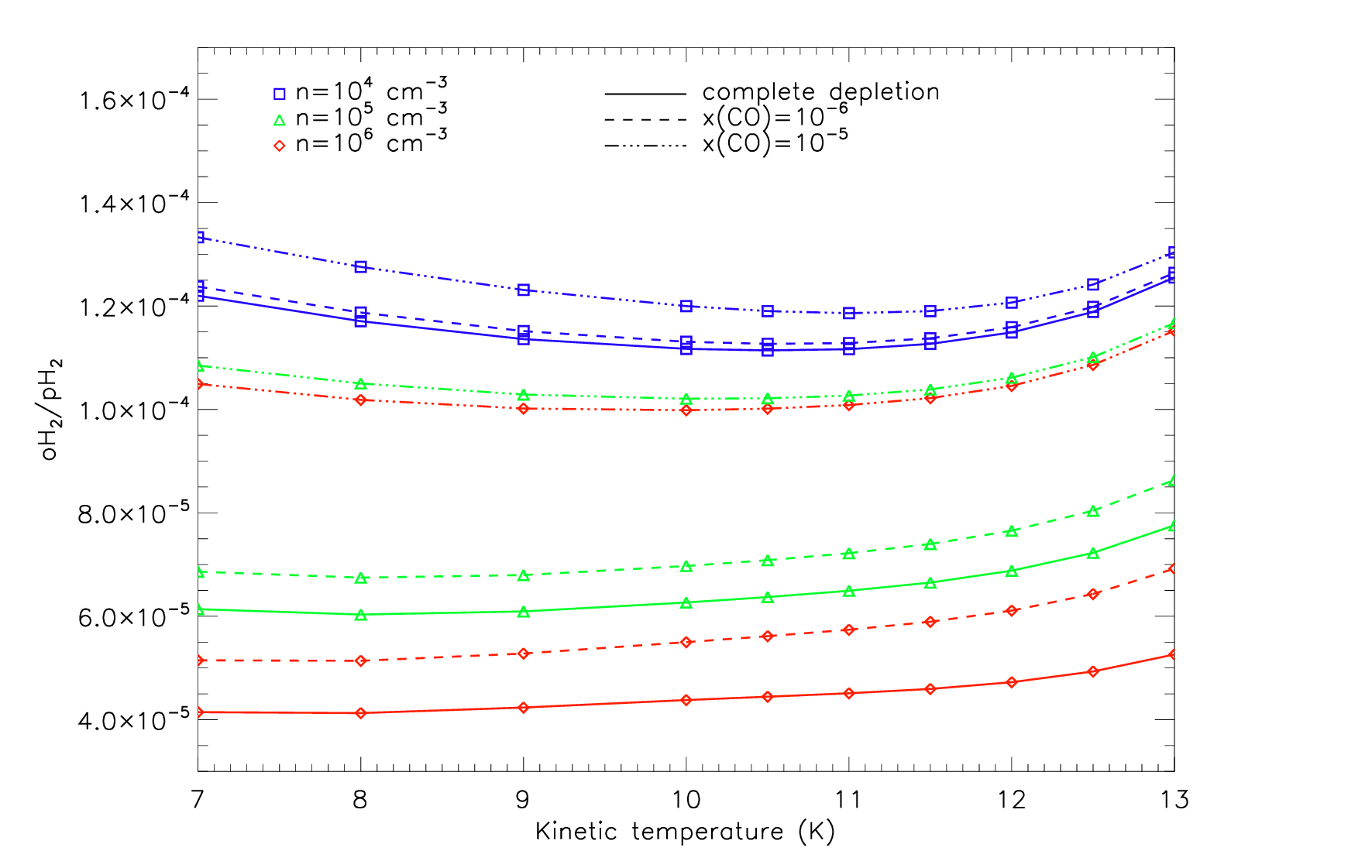}
\caption{Steady-state ortho-to-para ratio of H$_2$. }
\label{opH2}
\end{figure}

Because the rates for collisions with o-H$_2$ and p-H$_2$ are quite different, the nature of the collider matters for the excitation. 
Our chemical models (presented in Sect. \ref{chemicalmodel}) predict that the steady-state ortho/para-H$_2$ is within a factor of two around 10$^{-4}$ for all physical conditions probed (Fig. \ref{opH2}). We checked that the excitation is unchanged (within numerical precision) when considering only para-H$_2$ or para-H$_2$ with traces of ortho-H$_2$ (o/p-H$_2$\,$\le$\,2$\times$10$^{-4}$). We thus assumed in a first approximation collisions with para-H$_2$ only in the derivation of the column density of \ohhd and \pddh. \citet{Pagani09} however showed that the o/p-H$_2$ ratio is unlikely in steady-state in prestellar cores (see also Section \ref{timescales} and Fig. \ref{time_evol}). For completeness, we checked the computed column densities in the case where the o/p-H$_2$ ratio is higher than the steady-state value. 
We find that the derived column densities start to deviate sensibly from the ones derived with only p-H$_2$ for values of the o/p-H$_2$ ratio of the order of unity. 

The results with excitation with para-H$_2$ only are presented in Fig. \ref{coldens}a. The results with an o/p-H$_2$ ratio of 10$^{-2}$ are found to differ from Fig. \ref{coldens}a by less than 1\%. The column densities derived within the LTE approximation are also shown on the figure. Even at  a density of 10$^6$ cm$^{-3}$, the LTE assumption is not accurate, especially for \pddh which has the highest critical density. One can check indeed that the excitation temperature does not reach the kinetic temperature in any of the studied cases (Fig. \ref{coldens}b).

\begin{figure}[!h]
\includegraphics[width=9cm]{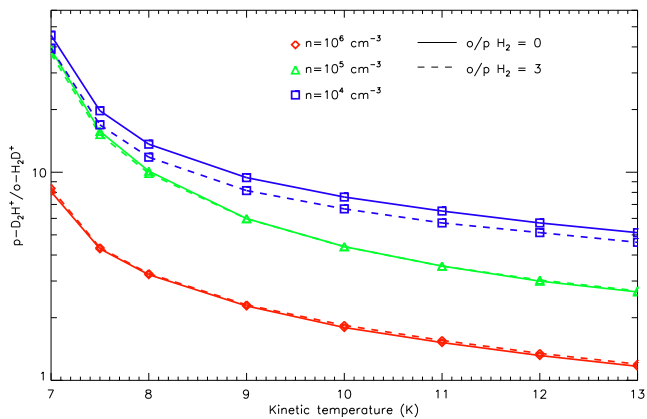}
\caption{\pddh / \ohhd abundance ratio resulting from the radiative transfer modelling of the observations, for different densities and o/p-H$_2$ ratio.}
\label{ratio}
\end{figure}

Assuming o/p-H$_2$ = 3 (the highest possible ratio at thermodynamical equilibrium) results in column densities 
about 10--20\% higher than in the case of excitation with only para-H$_2$, the effect being more important at low densities, as expected (at high densities, the system is close to LTE, and does not depend on the nature of the collider). The abundance ratio of the two molecules is even less sensitive to the o/p-H$_2$ ratio, as shown in Fig.  \ref{ratio}. This figure presents the \pddh / \ohhd~ratio resulting from our radiative transfer analysis, for different densities and in the two cases o/p-H$_2$ = 0 and 3. In the following, we aim at investigating the implications of this the \pddh / \ohhd~ratio with the help of a state-of-the art chemical model. 

\section{Astrochemical Modelling}
\label{chemicalmodel}

\subsection{Our model}

We solve the chemistry in the gas phase using a code developed by our group, and based on the DLSODES solver, which is part of the ODEPACK package (www.netlib.org). The code was benchmarked  against the results of \citet{Pagani09} and \citet{Sipila10}. We used the same approach as \citet{Sipila10} in building the reaction file for the complete depletion case. We take into account all different spin states of the different molecules and ions (i.e. we consider separately ortho and para species of H$_2$, D$_2$, H$_2$D$^+$, D$_2$H$^+$, H$_3^+$ and ortho, para and meta species of  D$_3^+$). 
We use the reaction rates between these species of \citet{Hugo09}. In addition we added a simple chemistry of CO and N$_2$, as in \citet{Pagani09}. We use the recombination rates for all ions of \citet{Pagani09}.

The pseudo-time-dependent chemical model contains only gas phase
reactions, except that formation of H$_2$, HD, and D$_2$ on the
grain surface is also included. We use the ``large grain limit''
\citep{Lipshtat04}, with the further assumption that all the
influx of H and D atoms is converted into H$_2$, HD, and D$_2$
molecules, and the diffusion rates of H and D atoms on the grain
surface are approximately the same. With these assumptions, the
formation rates of these three molecules can be written explicitly
as
\begin{eqnarray}
R {\rm (H_2)} &=& \frac{1}{2} n_{\rm G} \sigma_{\rm G} \frac{(V_{\rm
H}n_{\rm H})^2} {V_{\rm H}n_{\rm H}+V_{\rm D}n_{\rm D}}\\
R {\rm (HD)} &=& n_{\rm G} \sigma_{\rm G} \frac{V_{\rm H}n_{\rm
H}V_{\rm D}n_{\rm D}} {V_{\rm H}n_{\rm H}+V_{\rm D}n_{\rm D}}\\
R{\rm (D_2)} &=& \frac{1}{2} n_{\rm G} \sigma_{\rm G} \frac{(V_{\rm
D}n_{\rm D})^2} {V_{\rm H}n_{\rm H}+V_{\rm D}n_{\rm D}}
\end{eqnarray}
Here $n_{\rm G}$ and $\sigma_{\rm G}$ are the density and cross
section of grain particles, $V_{\rm H}$, $n_{\rm H}$, $V_{\rm D}$,
and $n_{\rm D}$ are the thermal velocity and number density of H and
D atoms. In this limit, the rather controversial detailed formation
mechanism of H$_2$ and the uncertain evaporation and diffusion
energy barriers do not enter the equations. The ortho and para forms
of H$_2$ and D$_2$ are produced according to the statistical weights
($3:1$ and $2:1$).

The influence of the different parameters on the chemistry has been studied in great detail by \citet{Walmsley04}, \citet{Flower04}, \citet{Pagani09} and \citet{Sipila10}. Here, we adopt a cosmic ray ionization of 3$\times$10$^{-17}$ s$^{-1}$, a dust-to-gas mass ratio of 1.3$\times$10$^{-2}$, and a grain radius of 0.1 $\mu$m. These parameters are not varied, because of the lack of observational constraints at hand. 

We simulate different levels of CO depletion by using different CO initial abundances. Note that we run the models with a constant elemental C and O abundance (i.e. we do not allow depletion on the grains during the evolution). In all our models, we assume that depletion affects CO and N$_2$ abundances at the same level.

\subsection{Timescales}
\label{timescales}

The free-fall timescale $\tau_{ff}$ of an isothermal sphere of $n$\,=\,10$^6$\,cm$^{-3}$ is 3$\times$10$^4$ yrs. 
Magnetic fields can increase the dynamical timescale, then controlled by ambipolar diffusion in the early evolution of subcritical cores.
Timescales of core evolution can be inferred empirically by comparing the number of prestellar and protostellar cores in observed regions \citep[for a review, see ][ and references therein]{Ward-Thompson07}. These empirical timescales lie roughly in the range 2--5 $\times$ 
$\tau_{ff}$.

\begin{figure}
\includegraphics[width=10.5cm]{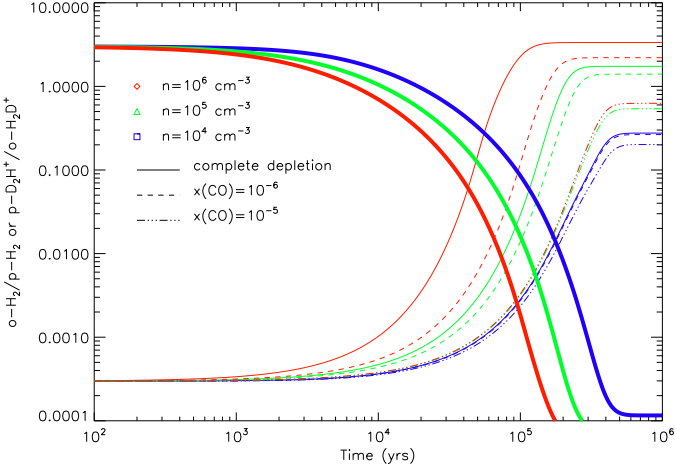}
\caption{Time evolution of the chemistry in the case of T\,=\,12\,K. The different colors stand for different densities: $n$=10$^4$ (blue), 10$^5$ (green), 10$^6$ cm $^{-3}$ (red). The thick lines represent the o/p H$_2$ ratio. The thin lines represent the 
p-D$_2$H$^+$/o-H$_2$D$^+$ ratio. The different line styles stand for different CO abundances: complete depletion (full), x$_{\rm CO}$\,=\,10$^{-6}$ (dash), x$_{\rm CO}$\,=\,10$^{-5}$ (dash-dot).}
\label{time_evol}
\end{figure}

It is instructive to compare the chemical timescales to this dynamical timescale. Figure \ref{time_evol} shows the time evolution of the o/p-H$_2$ ratio and p-D$_2$H$^+$/o-H$_2$D$^+$ ratio for constant physical conditions. The temperature was fixed to 12\,K, and three different densities were considered.
As already pointed out by e.g. \citet{Pagani09}, the longest timescale in the chemistry is the conversion between the two forms of ortho and para H$_2$. The other ions adjust to the o/p ratio of H$_2$ in typically less than 10$^2$\,yrs. The ortho-para interconversion of H$_2$ takes around 10$^4$\,yrs (typical timescale for decrease of o/p-H$_2$ ratio by a factor 1/e, see Fig.\,\ref{time_evol}). 

The free-fall timescale varies as 1/$\sqrt{n}$ and the chemical timescale varies even slower with the density. The  timescale for o/p-H$_2$ conversion is 6$\times$10$^3$ yrs at density $n$\,=\,10$^6$ cm$^{-3}$, and 9$\times$10$^3$ and 1.7$\times$10$^4$ yrs for densities of 10$^5$ and 10$^4$ cm$^{-3}$, respectively.
At high densities, the chemical timescales are somewhat shorter than the free-fall timescales (thus roughly one order of magnitude smaller than the empirical core timescales), and the difference even increases at lower densities. This comparison between timescales justifies at first order to consider the chemistry independently of the dynamical evolution, as already discussed by \citet{Walmsley04}.

Because the evolutionary stage of the H-MM1 core is not constrained, in the following we will focus on the study of the chemical composition when the steady-state is obtained. This choice avoids including the age of the core as an additional parameter but will not invalidate the generality of our conclusions. As shown in Fig. \ref{time_evol}, taking the chemical composition at long times overestimates the p-D$_2$H$^+$/o-H$_2$D$^+$ ratio in case the core is not yet in chemical steady-state. This note will be important for the extrapolation of our conclusions in the next sections.

\begin{figure}
\includegraphics[width=10.5cm]{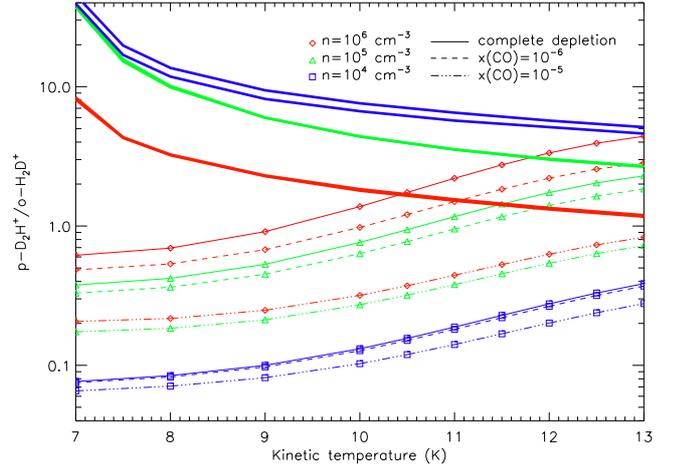}
\caption{Chemical model predictions of the p-D$_2$H$^+$/o-H$_2$D$^+$ ratio (same convention as previous figures), at steady-state. The thick decreasing lines show the ratio derived from the non-LTE analysis of the observations, in the assumption of different densities and different o/p-H$_2$ ratios (same curves as presented in Fig. \ref{ratio} but with the same linestyle for both o-p H$_2$ ratios).}
\label{obs_mod}
\end{figure}

\subsection{Physical conditions}

We could in principle derive the density profile of the core from the continuum emission at 850\,$\mu$m. However, this would require the knowledge of the temperature profile, which is not known. The Gould Belt Survey \citep{Andre10} made with the \textit{Herschel} Space Observatory should allow to derive accurate temperature profiles. Until then, as we cannot constrain the density profile without {\it ad-hoc} assumptions on the temperature,  
we restrict ourselves here in running chemical models at different densities, temperatures, and levels of CO depletion. Our goal is to investigate under which average conditions the model can reproduce the observations. 
A full physical model of the source will be presented in a forthcoming detailed study, after collecting new observational constraints.

Figure \ref{obs_mod} shows the prediction of the model for the p-D$_2$H$^+$/o-H$_2$D$^+$ ratio.
The thin lines are the model predictions, as a function of temperature, and for different densities and CO depletion factors. At fixed density, the p-D$_2$H$^+$/o-H$_2$D$^+$ ratio is increasing with the CO depletion level, because HD becomes more and more competitive over CO for reaction with H$_3^+$ and isotopologues. 
The ratio increases with temperature in this temperature range, as already explained in detail by \citet{Flower04}. 
The three thick lines are the ratios as calculated with RADEX from the observed line intensities (as a function of temperature and for different densities and o/p-H$_2$ ratios, as already shown in Fig. \ref{ratio}). We discuss in the following section the comparison between the chemical model and the observations.

\section{Discussion}

Figure \ref{obs_mod} allows to investigate the effect of the temperature, density and CO depletion level on the \pddh\,/\,\ohhd ratio, and to infer under which average conditions our model could reproduce the observed value. It should be noted that the observational ratio also depends on the assumed density, because of the non-LTE effects described in Section \ref{outLTE}. Agreement between observations and model can only be reached for densities strictly higher than 10$^5$\,cm$^{-3}$. This is consistent with the relatively high measured average density of the core \citep[5$\times$10$^5$ cm$^{-3}$,][]{Young06}. Even at the high density of 10$^6$\,cm$^{-3}$, the model can only reproduce the large observed p-D$_2$H$^+$/o-H$_2$D$^+$ ratio if $T>$10\,K and if the CO depletion level is substantial. The model points to the fact that the CO abundance should be $<$ 10$^{-5}$. This corresponds to a CO depletion level of more than 10 (for $T$\,=\,13\,K). The required depletion is even more severe if the density is a bit lower, or if the temperature is closer to 10\,K. Complete depletion (i.e. $>>$100) is required at 10\,K.  

\citet{Bacmann02} have studied the level of CO depletion toward a sample of seven prestellar cores. These  cores are located at distances between 120 and 200\,pc, and three of them are in the Ophiuchus molecular cloud. The cores have mean densities in the range (0.9--6)\,$\times$\,10$^5$\,cm$^{-3}$, similar to H-MM1. Based on IRAM observations of dust continuum and CO isotopologues (at angular resolution of 22$''$), the authors find depletion factors at the continuum peak of the cores in the range 4.5$-$15.5, but this factor drops to 2$-$5 at offset positions where the continuum is 70\% of the peak value. Although the study of Bacmann et al. (2002) was done at a 22$''$ resolution (to be compared to our 9--14$''$ beams), preventing us from doing an accurate comparison, the average depletion level we deduce in H-MM1 seems to be in the high range. Moreover the D$_2$H$^+$ emission is extended over 40$''$. Should the H$_2$D$^+$ be also extended on that scale, this would imply that the model requires a high CO depletion ($>>$ 10) over an extended region, in contradiction with the results of \citet{Bacmann02}. It is clear that the observation of the H$_2$D$^+$ spatial distribution and the direct measurement of the level of CO depletion are required to confirm if the astrochemical model is indeed unable to describe satisfactorily the roots of deuterium chemistry.

Following the discussion on the linewidths in Section \ref{linewidth}, the level of the thick observational curves in Fig. \ref{obs_mod} is in fact likely underestimated. Similarly, following the discussion in Section \ref{timescales}, the model predictions are overestimated if the core has not yet reached chemical equilibrium. These remarks show that the difficulties of the model might be even more severe than shown on Fig. \ref{obs_mod}.

The model also seems to exclude kinetic temperatures below 10\,K, but several very cold prestellar cores are known \citep[6--7\,K,][]{Crapsi07,Pagani07,Harju08}. The temperature in H-MM1 still needs to be determined, but this point could be another difficulty of the model.

Further tests of our understanding of the root of deuterium fractionation will depend on the observation of H$_2$D$^+$ at the offset positions where D$_2$H$^+$ was detected, and on the independent measure of the kinetic temperature and CO depletion level, for which we have an on-going APEX project.


\section{Conclusions}

We presented the first secure D$_2$H$^+$ detection in the ISM, toward the Oph H-MM1 core. The centroid velocity of the line is consistent with the one of the \ohhd~line. The emission is extended over 40$''$ at least. 
Using an out-of-equilibrium radiative transfer model, we have inferred the column densities of both \ohhd~and \pddh. We analyzed the chemistry with a state-of-the-art astrochemical model including all spin states of H$_3^+$, H$_2$D$^+$, D$_2$H$^+$, D$_3^+$. The model can reproduce the data only if the average density is high ($n$\,$>$\,few 10$^5$ cm $^{-3}$), the kinetic temperature is between 10 and 13\,K, and the depletion level of CO is high ($>$\,10 at 13\,K, $>$\,100 at 10\,K). Further observational studies may show whether these conditions indeed apply to H-MM1, and if the CO depletion level stays high on the large scales at which \pddh\ emits. 

Should new observational studies confirm the failure of the model, new ways will have to be investigated. We list here several possibilities:  include a more complete description of the chemistry (treating fully e.g. the CO chemistry and gas-grain interaction), recompute or measure the reaction and collision rates of all (spin separated) species, include in the chemical model a correct treatment of the excitation of the different molecules (at the moment, all adopted reactive and non-reactive collision rates describe the molecules in their rotational ground state).

\begin{acknowledgements}
      We thank the technical team who built the CHAMP$^+$ receiver for APEX.
      This work is supported by the German
      \emph{Deut\-sche For\-schungs\-ge\-mein\-schaft, DFG\/} Emmy Noether project
      number PA1692/1-1. 
\end{acknowledgements}

\bibliographystyle{aa}
\bibliography{/Users/bparise/These/Manuscrit/biblio}

\begin{thebibliography}{43}
\expandafter\ifx\csname natexlab\endcsname\relax\def\natexlab#1{#1}\fi

\bibitem[{{Amano} \& {Hirao}(2005)}]{Amano05}
{Amano}, T. \& {Hirao}, T. 2005, Journal of Molecular Spectroscopy, 233, 7

\bibitem[{{Andr{\'e}} {et~al.}(2010){Andr{\'e}}, {Men'shchikov}, {Bontemps},
  {K{\"o}nyves}, {Motte}, {Schneider}, {Didelon}, {Minier}, {Saraceno},
  {Ward-Thompson}, {di Francesco}, {White}, {Molinari}, {Testi}, {Abergel},
  {Griffin}, {Henning}, {Royer}, {Mer{\'{\i}}n}, {Vavrek}, {Attard},
  {Arzoumanian}, {Wilson}, {Ade}, {Aussel}, {Baluteau}, {Benedettini},
  {Bernard}, {Blommaert}, {Cambr{\'e}sy}, {Cox}, {di Giorgio}, {Hargrave},
  {Hennemann}, {Huang}, {Kirk}, {Krause}, {Launhardt}, {Leeks}, {Le Pennec},
  {Li}, {Martin}, {Maury}, {Olofsson}, {Omont}, {Peretto}, {Pezzuto}, {Prusti},
  {Roussel}, {Russeil}, {Sauvage}, {Sibthorpe}, {Sicilia-Aguilar}, {Spinoglio},
  {Waelkens}, {Woodcraft}, \& {Zavagno}}]{Andre10}
{Andr{\'e}}, P., {Men'shchikov}, A., {Bontemps}, S., {et~al.} 2010, \aap, 518,
  L102+

\bibitem[{Bacmann {et~al.}(2002)Bacmann, Lefloch, Ceccarelli, Castets,
  Steinacker, \& Loinard}]{Bacmann02}
Bacmann, A., Lefloch, B., Ceccarelli, C., {et~al.} 2002, A\&A, 389, L6

\bibitem[{{Belloche} {et~al.}(2006){Belloche}, {Parise}, {van der Tak},
  {Schilke}, {Leurini}, {G{\"u}sten}, \& {Nyman}}]{Belloche06}
{Belloche}, A., {Parise}, B., {van der Tak}, F.~F.~S., {et~al.} 2006, \aap,
  454, L51

\bibitem[{{Caselli} {et~al.}(2003){Caselli}, {van der Tak}, {Ceccarelli}, \&
  {Bacmann}}]{Caselli03}
{Caselli}, P., {van der Tak}, F.~F.~S., {Ceccarelli}, C., \& {Bacmann}, A.
  2003, \aap, 403, L37

\bibitem[{{Caselli} {et~al.}(2008){Caselli}, {Vastel}, {Ceccarelli}, {van der
  Tak}, {Crapsi}, \& {Bacmann}}]{Caselli08}
{Caselli}, P., {Vastel}, C., {Ceccarelli}, C., {et~al.} 2008, \aap, 492, 703

\bibitem[{{Crapsi} {et~al.}(2007){Crapsi}, {Caselli}, {Walmsley}, \&
  {Tafalla}}]{Crapsi07}
{Crapsi}, A., {Caselli}, P., {Walmsley}, M.~C., \& {Tafalla}, M. 2007, \aap,
  470, 221

\bibitem[{{Di Francesco} {et~al.}(2008){Di Francesco}, {Johnstone}, {Kirk},
  {MacKenzie}, \& {Ledwosinska}}]{diFrancesco08}
{Di Francesco}, J., {Johnstone}, D., {Kirk}, H., {MacKenzie}, T., \&
  {Ledwosinska}, E. 2008, \apjs, 175, 277

\bibitem[{{Dunham} {et~al.}(2008){Dunham}, {Crapsi}, {Evans}, {Bourke},
  {Huard}, {Myers}, \& {Kauffmann}}]{Dunham08}
{Dunham}, M.~M., {Crapsi}, A., {Evans}, II, N.~J., {et~al.} 2008, \apjs, 179,
  249

\bibitem[{{Emprechtinger} {et~al.}(2009){Emprechtinger}, {Caselli}, {Volgenau},
  {Stutzki}, \& {Wiedner}}]{Emprechtinger09}
{Emprechtinger}, M., {Caselli}, P., {Volgenau}, N.~H., {Stutzki}, J., \&
  {Wiedner}, M.~C. 2009, \aap, 493, 89

\bibitem[{{Flower} {et~al.}(2004){Flower}, {Pineau des For{\^e}ts}, \&
  {Walmsley}}]{Flower04}
{Flower}, D.~R., {Pineau des For{\^e}ts}, G., \& {Walmsley}, C.~M. 2004, \aap,
  427, 887

\bibitem[{{G{\"u}sten} {et~al.}(2006){G{\"u}sten}, {Booth}, {Cesarsky},
  {Menten}, {Agurto}, {Anciaux}, {Azagra}, {Belitsky}, {Belloche}, {Bergman},
  {De Breuck}, {Comito}, {Dumke}, {Duran}, {Esch}, {Fluxa}, {Greve}, {Hafok},
  {H{\"a}upl}, {Helldner}, {Henseler}, {Heyminck}, {Johansson}, {Kasemann},
  {Klein}, {Korn}, {Kreysa}, {Kurz}, {Lapkin}, {Leurini}, {Lis}, {Lundgren},
  {Mac-Auliffe}, {Martinez}, {Melnick}, {Morris}, {Muders}, {Nyman}, {Olberg},
  {Olivares}, {Pantaleev}, {Patel}, {Pausch}, {Philipp}, {Philipps},
  {Sridharan}, {Polehampton}, {Reveret}, {Risacher}, {Roa}, {Sauer}, {Schilke},
  {Santana}, {Schneider}, {Sepulveda}, {Siringo}, {Spyromilio}, {Stenvers},
  {van der Tak}, {Torres}, {Vanzi}, {Vassilev}, {Weiss}, {Willmeroth},
  {Wunsch}, \& {Wyrowski}}]{Guesten06SPIE}
{G{\"u}sten}, R., {Booth}, R.~S., {Cesarsky}, C., {et~al.} 2006, in Society of
  Photo-Optical Instrumentation Engineers (SPIE) Conference Series, Vol. 6267,
  Society of Photo-Optical Instrumentation Engineers (SPIE) Conference Series

\bibitem[{{Harju} {et~al.}(2008){Harju}, {Juvela}, {Schlemmer}, {Haikala},
  {Lehtinen}, \& {Mattila}}]{Harju08}
{Harju}, J., {Juvela}, M., {Schlemmer}, S., {et~al.} 2008, \aap, 482, 535

\bibitem[{Hirao \& Amano(2003)}]{Hirao03}
Hirao, T. \& Amano, T. 2003, ApJ, 597, L85

\bibitem[{{Hugo} {et~al.}(2009){Hugo}, {Asvany}, \& {Schlemmer}}]{Hugo09}
{Hugo}, E., {Asvany}, O., \& {Schlemmer}, S. 2009, J. Chem. Phys., 130, 164302

\bibitem[{{Johnstone} {et~al.}(2004){Johnstone}, {Di Francesco}, \&
  {Kirk}}]{Johnstone04}
{Johnstone}, D., {Di Francesco}, J., \& {Kirk}, H. 2004, \apjl, 611, L45

\bibitem[{{J{\o}rgensen} {et~al.}(2008){J{\o}rgensen}, {Johnstone}, {Kirk},
  {Myers}, {Allen}, \& {Shirley}}]{Jorgensen08}
{J{\o}rgensen}, J.~K., {Johnstone}, D., {Kirk}, H., {et~al.} 2008, \apj, 683,
  822

\bibitem[{{Khanzadyan} {et~al.}(2004){Khanzadyan}, {Gredel}, {Smith}, \&
  {Stanke}}]{Khanzadyan04}
{Khanzadyan}, T., {Gredel}, R., {Smith}, M.~D., \& {Stanke}, T. 2004, \aap,
  426, 171

\bibitem[{{Klein} {et~al.}(2006){Klein}, {Philipp}, {Kr{\"a}mer}, {Kasemann},
  {G{\"u}sten}, \& {Menten}}]{Klein06}
{Klein}, B., {Philipp}, S.~D., {Kr{\"a}mer}, I., {et~al.} 2006, \aap, 454, L29

\bibitem[{{Lipshtat} {et~al.}(2004){Lipshtat}, {Biham}, \&
  {Herbst}}]{Lipshtat04}
{Lipshtat}, A., {Biham}, O., \& {Herbst}, E. 2004, \mnras, 348, 1055

\bibitem[{{Loinard} {et~al.}(2008){Loinard}, {Torres}, {Mioduszewski}, \&
  {Rodr{\'{\i}}guez}}]{Loinard08}
{Loinard}, L., {Torres}, R.~M., {Mioduszewski}, A.~J., \& {Rodr{\'{\i}}guez},
  L.~F. 2008, \apjl, 675, L29

\bibitem[{{Loren} {et~al.}(1990){Loren}, {Wootten}, \& {Wilking}}]{Loren90}
{Loren}, R.~B., {Wootten}, A., \& {Wilking}, B.~A. 1990, \apj, 365, 269

\bibitem[{{Motte} {et~al.}(1998){Motte}, {Andre}, \& {Neri}}]{Motte98}
{Motte}, F., {Andre}, P., \& {Neri}, R. 1998, \aap, 336, 150

\bibitem[{{M{\"u}ller} {et~al.}(2005){M{\"u}ller}, {Schl{\"o}der}, {Stutzki},
  {Schlemmer}, {Giesen}, \& {Schilke}}]{Muller05}
{M{\"u}ller}, H.~S.~P., {Schl{\"o}der}, F., {Stutzki}, J., {et~al.} 2005, in
  IAU Symposium, ed. D.~C. {Lis}, G.~A. {Blake}, \& E.~{Herbst}, 30--+

\bibitem[{{Pagani} {et~al.}(2007){Pagani}, {Bacmann}, {Cabrit}, \&
  {Vastel}}]{Pagani07}
{Pagani}, L., {Bacmann}, A., {Cabrit}, S., \& {Vastel}, C. 2007, \aap, 467, 179

\bibitem[{{Pagani} {et~al.}(2009){Pagani}, {Vastel}, {Hugo}, {Kokoouline},
  {Greene}, {Bacmann}, {Bayet}, {Ceccarelli}, {Peng}, \&
  {Schlemmer}}]{Pagani09}
{Pagani}, L., {Vastel}, C., {Hugo}, E., {et~al.} 2009, \aap, 494, 623

\bibitem[{{Parise} {et~al.}(2004){Parise}, {Castets}, {Herbst}, {Caux},
  {Ceccarelli}, {Mukhopadhyay}, \& {Tielens}}]{Parise04}
{Parise}, B., {Castets}, A., {Herbst}, E., {et~al.} 2004, \aap, 416, 159

\bibitem[{{Parise} {et~al.}(2002){Parise}, {Ceccarelli}, {Tielens}, {Herbst},
  {Lefloch}, {Caux}, {Castets}, {Mukhopadhyay}, {Pagani}, \&
  {Loinard}}]{Parise02}
{Parise}, B., {Ceccarelli}, C., {Tielens}, A.~G.~G.~M., {et~al.} 2002, \aap,
  393, L49

\bibitem[{{Parise} {et~al.}(2009){Parise}, {Leurini}, {Schilke}, {Roueff},
  {Thorwirth}, \& {Lis}}]{Parise09}
{Parise}, B., {Leurini}, S., {Schilke}, P., {et~al.} 2009, \aap, 508, 737

\bibitem[{{Ramanlal} \& {Tennyson}(2004)}]{Ramanlal04}
{Ramanlal}, J. \& {Tennyson}, J. 2004, \mnras, 354, 161

\bibitem[{{Risacher} {et~al.}(2006){Risacher}, {Vassilev}, {Monje}, {Lapkin},
  {Belitsky}, {Pavolotsky}, {Pantaleev}, {Bergman}, {Ferm}, {Sundin},
  {Svensson}, {Fredrixon}, {Meledin}, {Gunnarsson}, {Hagstr{\"o}m},
  {Johansson}, {Olberg}, {Booth}, {Olofsson}, \& {Nyman}}]{Risacher06}
{Risacher}, C., {Vassilev}, V., {Monje}, R., {et~al.} 2006, \aap, 454, L17

\bibitem[{{Roberts} {et~al.}(2003){Roberts}, {Herbst}, \& {Millar}}]{Roberts03}
{Roberts}, H., {Herbst}, E., \& {Millar}, T.~J. 2003, \apjl, 591, L41

\bibitem[{{Roueff} {et~al.}(2007){Roueff}, {Parise}, \& {Herbst}}]{Roueff07}
{Roueff}, E., {Parise}, B., \& {Herbst}, E. 2007, \aap, 464, 245

\bibitem[{{Sipil{\"a}} {et~al.}(2010){Sipil{\"a}}, {Hugo}, {Harju}, {Asvany},
  {Juvela}, \& {Schlemmer}}]{Sipila10}
{Sipil{\"a}}, O., {Hugo}, E., {Harju}, J., {et~al.} 2010, \aap, 509, A98

\bibitem[{{Stanke} {et~al.}(2006){Stanke}, {Smith}, {Gredel}, \&
  {Khanzadyan}}]{Stanke06}
{Stanke}, T., {Smith}, M.~D., {Gredel}, R., \& {Khanzadyan}, T. 2006, \aap,
  447, 609

\bibitem[{Turner(1990)}]{Turner90}
Turner, B.~E. 1990, ApJ, 362, L29

\bibitem[{{Turner}(2001)}]{Turner01}
{Turner}, B.~E. 2001, \apjs, 136, 579

\bibitem[{{van der Tak} {et~al.}(2007){van der Tak}, {Black}, {Sch{\"o}ier},
  {Jansen}, \& {van Dishoeck}}]{vanderTak07}
{van der Tak}, F.~F.~S., {Black}, J.~H., {Sch{\"o}ier}, F.~L., {Jansen}, D.~J.,
  \& {van Dishoeck}, E.~F. 2007, \aap, 468, 627

\bibitem[{{van der Tak} {et~al.}(2005){van der Tak}, {Caselli}, \&
  {Ceccarelli}}]{vanderTak05}
{van der Tak}, F.~F.~S., {Caselli}, P., \& {Ceccarelli}, C. 2005, \aap, 439,
  195

\bibitem[{{Vastel} {et~al.}(2004){Vastel}, {Phillips}, \& {Yoshida}}]{Vastel04}
{Vastel}, C., {Phillips}, T.~G., \& {Yoshida}, H. 2004, \apjl, 606, L127

\bibitem[{{Walmsley} {et~al.}(2004){Walmsley}, {Flower}, \& {Pineau des For{\^
  e}ts}}]{Walmsley04}
{Walmsley}, C.~M., {Flower}, D.~R., \& {Pineau des For{\^ e}ts}, G. 2004, \aap,
  418, 1035

\bibitem[{{Ward-Thompson} {et~al.}(2007){Ward-Thompson}, {Andr{\'e}},
  {Crutcher}, {Johnstone}, {Onishi}, \& {Wilson}}]{Ward-Thompson07}
{Ward-Thompson}, D., {Andr{\'e}}, P., {Crutcher}, R., {et~al.} 2007, Protostars
  and Planets V, 33

\bibitem[{{Young} {et~al.}(2006){Young}, {Enoch}, {Evans}, {Glenn}, {Sargent},
  {Huard}, {Aguirre}, {Golwala}, {Haig}, {Harvey}, {Laurent}, {Mauskopf}, \&
  {Sayers}}]{Young06}
{Young}, K.~E., {Enoch}, M.~L., {Evans}, II, N.~J., {et~al.} 2006, \apj, 644,
  326

\end{thebibliography}

\end{document}